# Implementation of Electrical Feedback Technologies in 5 Households in Ankara, Turkey

Hatice Şengül [a], Tahsin Olgu Benli [b]

[a] Environmental Engineering Department, Hacettepe University, P.O. Box 06800 Beytepe, Ankara, Turkey

[b] Graduate School of Clean and Renewable Energies, Hacettepe University, P.O. Box 06800 Beytepe, Ankara, Turkey

**Abstract:** Most people don't know how much energy they are spending for different purposes and also they are unaware of potential electrical consumption reduction level they could make by changing their consumption behaviors or investing in new efficient technology products. Providing electrical consumption information to household and make the consumption information tangible is very effective strategy in reducing electrical consumption in households. In-Home Displays (IHDs) provide consumers real time information on electrical consumption level and related expenditure, with that strategy turning an intangible electric bill into transparent and controllable is possible. We interviewed the participants before the start to our study in order to have knowledge of their income, environmental attitudes and the knowledge level of their electrical consumption conserving activities. Monitoring over 3 to 5 months maximal of five households' electrical consumption and comparing that consumption level with the previous year had been realized. Households received advanced real time appliance level feedback. Behavioral change associated electrical consumption reduction levels were identified. It is also realized that setting goals via an electrical energy consumption display have high potential on reducing the total electricity consumption in households. We found 0 %, 0 %, 13.21 %, 3.42 %, 1.59 % electricity savings on house 1, house 2, house 3, house 4 and house 5 correspondingly. It is suggested that further attention need to be given on potential reduction effect of applying different type of monitor designs and implication of special designed stimulants that can be used all over the house that can alert households with their current consumption level. Also, it is suggested to concentrate on time periods in order to analyze and try to reveal specific time periods for specific consumption behaviors.

**Key words:** Feedback technologies, in home displays, electrical consumption reduction, behavioral change of consumers, monitoring of electrical usage in households, appliance level feedback, real time feedback

## 1. Introduction

Mankind now is facing with the global environmental problems that have threaten their lives. Excessive use of fossil fuels such as coal and oil cause environmental problems like acid rains and greenhouse effect. It is known that global carbondioxide emissions grow rapidly, if the necessity precautions on climate change won't taken, the total carbondioxide emission amount will be doubled in 2050 [1]. Energy demand of countries increasing, due to that situation external energy sources dependence also increase. Therefore, countries have difficulties with balancing the supply demand profiles and also they become indebted to external energy sources continuously.

It is accepted that sustainability level of energy sources due to high volume of energy consumption in the industralized nations is scanty, mainstream is the mankind hardly continue to maintain the desired necessity energy sources level. The problems caused by excessive energy consumption levels, and the solution may be come in two major ways: fronting to new energy sources or altering the energy consumptions. In the short time period, second solution seems to be feasible. In brief, we have to develop energy efficient approaches, and make energy efficient replacements in our daily lives.

According to international energy agency; energy effiency is the key strategy in the sustainable growth process. Increasing energy demand of housings cause residentials to become one of the major consumers in the energy supply demand balance. In accordance with the some estimates, in near future, residential annual energy consumption amount will correspond to 40% of total annual energy consumption of world [2]. When merely take the total electricity consumption into account, the total electricity consumption on dwellings correspond to 25% of the total electricity consumption in residential sector [3].

As we stated before carbon dioxide is the most important greenhouse gas. The using of fossil fuels such as coil and oil induced the greenhouse effects and acid rains. With the fact that the effects of climate change harm the world, high energy prices, increase at the supply demand, improving the energy efficiency on residential sector, fronting to use of energy efficient new technology products and using of renewable energy sources have to be stimutaled. In 2002, the total emission amount which were caused by the carbon dioxide, corresponded to %82 of the total emission amount in the Europe. %39 of the total carbon dioxed based emission occured from the use of electricity and heating productions. Buildings are the major pollution sources. The four biggest source that caused to form greenhouse gas in european union was the buildings originating emissions [4].

The total energy that consumed in the existing building stock in the countries which are the members of the european union equal to 40% of the total energy consumption that was consumed by the all the european union member countries. The 63% of the total energy consumption of building stock comprised of residential sector's total energy use amount. If we merely take the electrical consumption amount into account, the electrical energy amount which consumed by the building stock corresponds to 73% of the total consumption amount. For this reason, improving the total energy performance of the residential sector cause to reducing the external dependence of the countries and in addition that strategy will make possible of catching up with the carbon dioxide emission levels that had been agreed at the Kyoto Procotols [5].

With the surface area and the population, Turkey is remarkable country in the world. The population exceeds 74.7 million. According to 2011 datas, gross national product equaled to 772,298 billion dolar and per capita income equaled to 10444 $. The total energy consumption of Turkey in 2011 accounted to 118,8 million tons of oil equivalent. The total electricity consumption in 2011, accounted to 224,41 billion kWh with the %8 increase due to previous year [6].

The primal energy consumption amount ratio to total national product equals to energy density. The countries' development level are defined with the electrical consumption amount per person and also with the energy density. High electricity consumption amount per person is the sign of the high level of economic development and wealth level of that country. The lowness in the energy density means more work with the same energy. The countries with the high ratio of per person electrical consumption amount are: Iceland, Norway, Kuwait, Qatar, Canada, Sweden and United States. Among these countries, Norway, Sweden and United States have lower level of energy densities, therefore we can say the wealth level of that countries are higher than the others.

In the year 2011, per person energy consumption in the world accounted for 1.87 tons of oil equivalent and per person electrical consumption in the world eqauled to 3155 kWh. In the same year, with the 1.59 tons of oil equivalent per person and with the 3058 kWh total electrical consumption amount Turkey found place on the middle. The Turkey's goal for the per person electrical consumption amount for the 2015 year was 3600-3800 kWh, for the year

2020 is 4800-5000 kWh, for the year 2030 is more than 7000 kWh and for the year 2040 is more than 8000 kWh [7].

The greenhouse emission of Turkey had increased significantly. 170 million tons of carbon dioxide equivalent in the year 1990 increased to 366.5 million tons of carbon dioxide equivalent in 2008. Per person carbon dioxide amount accounted to 5.5 kg. According to United Nation Climate Change Framework Convention, between the countries that were listed in appendix-1, Turkey was the country which had the worst greenhouse gas emission increase rate.

In accordance with the 2008 year Greenhouse Gases National Stock declaration, 16% of Turkey's national carbon dioxide amount and 18% of the energy sectror comprised of the total building stocks in Turkey. According to existing case scenario, 28.3 million tons of oil equivalent energy consumption will assumed to rise 47.5 million tons of oil equivalent in 2020, that corresponds to double amount of existing carbon dioxide emission amount. On the other hand, it shouldn't forgotten that there is lot of potential in the building stock for the taking the energy efficient decisions and implementing [8].

Turkey import 75% of the total energy that is been used. Supplying the major amount of energy demand from importing threats Turkey's energy safety. Accounting to projection which had been prepared by ministry of energy and national resources of Turkey, the primal energy demand of Turkey assumed to raise from 170 million ( equivalent amount in 2015) to 222 million tons of oil equivalent in 2020. Also it is assumed that when supplying the total energy demand Turkey will import 70% of the total energy need in 2020. That procejtions indicated that without taking necessity precautions it is clearly obvious that solving the Turkey's external dependence problem is infeasible [9].

External dependence on energy also jeopardise the balance of the electrical energy demand. The gross electricity demand of Turkey in 2011 equaled to 230.3 billion kWh with the 9.4% increase ratio to previous year. Net electricity consumption of the Turkey equaled to 186 billion kWh in 2011. The total electrical energy production in Turkey for the year 2012 equaled to 239 billion kWh, meantime the consumption amount equaled to 241 billion kWh. Accourding to 2013 year electrical production datas of Turkey showed that 239 billion kWh electricity was produced while the consumption amount equaled to 245 billion kWh [10]. The results showed that Turkey has to put energy efficient strategies into practice.

One another projection which predicted the total electricity demand amount with the time period of 2012 – 2021 was done by ministry of energy and natural resources of Turkey. The electricity demand in 2012 was 244026 GWh, in 2013 was 262010 GWh, in 2014 was 281859 GWh. In 2016, it was predicted that the total electricity will increase with the 7.6% increase rate comparing to previous year, for the 2017 the total electricity demand will increase to 350300 GWh with the 7.5% increase rate comparing to 2016, the total electricity demand will be equal to 404160 GWh for the year of 2019, it will account to 433900 GWh with the 7.4% increase rate comparing to previous year and in the year 2021 we will see 7.7% increase rate and it will correspond to total amount of 467260 GWh [11].

Electricity consumption in the residential sector have been politically and scientifically the center of interest. Balancing the supply demand, controlling the demand of electrical need and implication of renewable energy systems into existing power system have been desired. In accordance with this purpose, manifesting the households' electrical consumption relationship with the consumers' consumption behaviors need to be clarified.

Giving the necessity knowledge of information that ensure savings, changing of consumption behaviors into effective way that cause savings have to be implemented. Persistent measurements for the electrical consumption amounts and the analysis on the total consumption has to be done in order to let the consumer to invest new energy efficient technologies that have less electricity consumption, thus changing of the existing consumer behaviors can be altered in a positive way. Usually, the profit that can be provided by changing the old technology product with the new one is far greater than the monetary value that will be needed to invest when buying new energy efficient technology products [12]. With that strategy, consumers can also alter when they will use the products, for instance they probably use the major electrical consumption devices in the cheaper time period which is indicated in their electrical bills also they can invest in renewable energy applications in order to supply some of the electricity demand from there [13]. That strategy is crucial because by the help of that approach, there will be high potential emerge with the reduction in the electricity and carbon dioxide use amount.

It is preferred buy cheaper and less energy efficient technological products, due to higher efficiency equipments have higher prices [14]. Also when we look from the point of the view of manufacturers, they want to sell more products so their approach is sell cheaper products. Energy labels of appliances are helping customers with gaining knowledge in energy performace at the point of sale. However as we stated before when decision has been made to buy new appliance, the regret won't be occured until thought of investing in new less energy consumption devices has been emerged. For those reasons, investing new technology approach also have to be reinforced with another strategy, changing the consumption behaviors.

Prominent strategy is to analyse and get the consumption amounts of major electrical consumers appliance individually, with that approach explicit knowledge of net consumption patterns can be lightened. We need feedback system which will be connected to main fuse in the houses which can gauge total electrical consumption of house in real time, also if we need to get the consumption patterns of individuals we have to set up the disaggregated feedback system which have the capability of gauging electrical consumptions of individuals. Monetary savings alternatives can be suggested after the gauging the electrical consumption patterns of houses. It is also need to be clarified that which consumption behavior seems easy to change for savings goal and which seems tough to change due to daily life activities.

Feedback technology products need display which give the information about the rates of electrical energy which the consumers use in their own homes for promoting energy saving actions. The prices of digital in home displays are affordable, still in home displays are need to be designed to have maximum efficient stimuli effect that can let consumers to change their behavior. Besides technological features of in home displays, also there are other factors that have effect on electricity savings: exact location of in home display and also whether it should be fixed or portable, whether beside having small in home display also is there any need to bigger lcd display in which consumers can realize their real time and also historic consumption behavior via graphical analysis. With the previous studies, it had been indicated that the most effective type of feedback system's display should be chosen with small in home display in conjuction with bigger central lcd display [15].

It is reasonable that if the occupants have chance to know information about specific electrical energy consumption behaviors, they intend to change their consumption behaviors. By potential future monetary savings, they maybe want to spend their moneys into another activities or investments. Various researches have studied the effects of feedback on residential electricity consumption and they found that feedback can be effective way of reducing the electricity consumption whether by applying alone or in conjuction with other factors.

## 2. Literature

Feedback concept can be defined as; evaluable or recoverable transmission process of movement, action, process or sources that have been controlled. Research based upon residential areas we can use the definition of total electricity consumption of individual households, for the feedback concept [16]. Classification of houses' electricity consumption information hasn't been accepted as a research area that has been recently emerged. Frequency and the level of feedback information projections' had been researched in the 1970's and 1980's. Supplying the total consumption information to consumers via electric bill doesn't suffice for realizing the net relationship between total electricity consumption amount and the related monetary value of that consumption. Thus, electric bill can't prevent the waste use of electrical energy. In the researches that had been done; among the various feedback information, question of which information is the most important determinant that cause reduction in the total electricity consumption had been analyzed.

In 1990's new informing mechanism emerged. With the use of energy efficient gauges it was targeted to gathering detailed data of consumers' electricity consumptions by the providing consumers to total control on the total electricity usage. With all those information developing detailed characterization over consumers' electricity consumptions had been targeted. In the latest 7-10 years, with the help of the continuously developing web based activities, processing of gathered data is now easier. Advanced counters have also been supplied to consumers by the providers. Those devices can store the electrical consumption information in their memories and also can display that specific information in the display unit in the houses. With computers or another devices, it is also possible to make further analysis on that consumption data.

By the help of the pilot studies which had been executed in the latest 7-8 years, effect of the dynamic feedback technologies over occupants had been investigated, been characterized and been elucidated. With the help of those pilot studies significant progress has been made, however it is not been clarified that in which order that feedback technologies will induce the behavior of consumers and whether those effects will be permanent or temporary.

### 2.1 General Overview to Feedback Effects

Electricity consumption savings effects of the feedback technologies have been examined under two disciplines: behavior science and economical aspect.

### 2.1.1    Economical Effect of Feedback Technologies

When the consumers don't know the related net electrical consumption amount of individual appliances, they can't catch up with the consumption plan and they will probably consume more or less than this amount. Feedback technologies let the consumers keep the balance with the monetary value of electric bills and budget they allocated. With the use of feedback technologies in households the uncertainty of how much money the consumers will pay for the specific usage will be resolved. With disambiguation, average electricity consumption can be reduced and the savings can also be altered into other usages. Adjustments on the existing electricity consuming device's usage time can also provide electricity savings. Over the long run consumers can prefer to replace old existing electricity consumption devices with the new energy efficient ones.

### 2.1.2 Behavioral Effect of Feedback Technologies

Investment based behaviors are the non-repeated behaviors that based upon buying new energy efficient devices for once. Day to day behaviors is defined as the behaviors that exhibited less frequently, also called as naturally repeated behaviors (Barr et. al, 2005). As an example of effect on the consumer's behaviors, reduction of the electricity usage amount in the peak electrical tariff period by the switching the usage time into cheaper electrical tariff period can be accepted as one of the behavioral effect of feedback technologies. There are some strategies for changing the consumers' behaviors, most known two strategies are; foregoer strategy and outcome strategy. In foregoer strategies, changing of the behaviors while the actions are taking place is aimed, generally they are not personal and they target big populations. In the outcome strategies, changing the consumers' behaviors by the method of informing the consumers with related consumption behaviors' monetary value, is aimed. Rewarding strategies which prevent some specific consumption behaviors and mission strategies which are based on goal settings are mostly used strategies in the behavior concept. Also it is possible to motivate the occupants with the intention of gaining money, when the electricity savings will occur, some monetary rewards can be given.

### 2.1.3 Feedback Technologies' Prosperousness Conditions

Feedback technologies should cause changes in savings supportive behaviors. In one definition; " if the real time consumption is lesser than the consumption goal level, it shows us the feedback technologies system works well for that example " (Selingman et. al, 1981). The most important part in the feedback studies is setting the goal and trying to catch up with that goal. In another study, it was said that we could accept the feedback technologies were successful if three stages were realized; stage of learning, stage of gaining the habit and stage of internalization of behaviors (Raaij and Verhallen 1983). In the learning stage consumers have chance to get knowledge of their specific consumption behavior profiles. If the feedback information display frequency is fast enough, consumers can realize how order their specific consumption behaviors effect the total consumption amount. They can make electricity saving changes in their consumption behaviors with the purpose of reducing the total consumption amount. If the consumers will maintain behaviors that can let the reduction in the electricity consumption amount for a length of time, these behaviors will become their new habits. It was indicated that the time need for changing of behaviors into new habits is minimum three months and after that three months feedback technology devices shouldn't be removed from houses (Darby 2006). Stage of internalization of behaviors is the third stage that is needed for prosperousness conditions for feedback technologies. When consumers are turning their new behaviors into habits, their approaches also should be develop in a way of adjusting the existing behaviors. In other studies it was said that removing of various behavior changing strategies; such as rewarding with monetary value or sending supportive messages, from the application area after some period of time will let the electricity saving effect will be disappeared (Katdev and Johnson 1987).

## 2.2 Feedback Technologies Classification Method

Classification of feedback technologies that will be used in the researches can be categorized by; types of feedback, frequency of feedback information, working principal of information sending mechanism and the way of electricity cost information features of feedback technology products. In Darby classification spectrum, we have two different segmentations: direct feedback and indirect feedback. The choose criterion to whether use direct or indirect feedback technology product is determined by looking to electricity consumption projection to consumer feature of the product, if the feedback technology device has ability to transmit the total electricity consumption to consumers, we said that type of feedback is listed under the direct feedback, contrary if the feedback technology doesn't provide projection of the electrical consumption amount to consumers it is going to be listed in indirect feedback technologies

(Darby 2000).

### 2.2.1 Indirect Feedback Mechanisms

In indirect feedback mechanisms, feedback of consumption information will be transmitted to consumers after the consumption has been made. There are four different indirect feedback mechanism classifications: standard type billing, more informative billing system, foreseen billing system and daily or weekly feedback method.

#### 2.2.1.1 Standard Type Billing

In this method, the bills will be presented to customers at the end of the months. In bills, customers can find total kWh amounts and corresponding consumption pricing information. General use area of that type of indirect feedback is to show total consumption monetary value.

#### 2.2.1.2 More informative Billing

In this method, more detailed total electricity consumption amount information of households is presented. Consumers can find comparative statistics about their consumption behaviors. Typically, recent total electricity consumption amount and total kWh amount information are compared with previous months.

#### 2.2.1.3 Foreseen Billing System

Foreseen billing system is a bottom up forecasting method, which uses the information of types of customers, appliances that are been used in the houses and type of billing in statistics techniques to forecast the total electricity consumption amount. With that method, we can merely see forecasted results for the total electricity consumption but we couldn't see individual appliance consumption.

#### 2.2.1.4 Daily or Weekly Feedback Method

In that kind of indirect feedback method, consumers want from suppliers to provide daily or weekly total electricity consumption amount. Also, they can note down the total daily electricity consumption by checking their own electrical counter and they make the analysis by their selves.

### 2.2.2 Daily or Weekly Feedback Method

In direct feedback mechanisms, transmitting of total electricity consumption amount and the corresponding monetary value will be realized in real time during the consumption process. This category is divided into two sub categories: real time direct feedback method and real time advanced feedback method.

#### 2.2.2.1 Real Time Direct Feedback Method

The most used direct feedback technology products are the in home displays that are used in houses and show the total real time electricity consumption to the occupants. Those devices read the electrical consumption value from the house's main electrical switch board panel or from main electrical counter and those devices let this information to be showed in display in real time with the purpose of sharing the total electricity consumption information with occupants. With showing the real time electricity consumption via display occupants have chance to assess specific electricity consumption behaviors over total consumption.

#### 2.2.2.2 Real Time Advanced Direct Feedback Method

In that type of feedback along with the showing the total electricity consumption, consumers also can see the each individual appliances' real time total electricity consumptions. Most of the time, technological feedback products which belong to that category also provide remote control of each individual appliance like open and close commands.

One of the most important feature for that category is having the user interface. In the user interface which is mostly serviced by websites of related products, consumer can see and analyze each individual appliance's electricity consumption amounts. They can also see and compare the historical consumption amounts and corresponding monetary value on the daily, weekly or monthly basis.

## 2.3 Saving Results of Some Feedback Studies in the Literature

We want to give some results of studies that had been completed under two categories: total savings when behavioral and psychological effect can occurred with feedback technologies products and the second category referred as total savings due to technological and methodical variations.

### 2.3.1 Saving Results Due to Technological and Methodical Variations

By the use of real time direct feedback method in 12 houses with the duration of six months, the average electricity saving was calculated as 7 % (Kirkeide, 2009). By using real time advanced direct feedback method in 5 houses with the sampling interval between 4 – 25, average electricity savings of 12 % was found (Darby 2006). With the intention of investigating the effect of self-monitoring to electricity savings, 4 – 6 weeks studies were conducted in Washington and the results indicated that overall electricity savings of 11 % (Winett et. al, 1979). In Kyoto, Japan a monitoring project in order to show end use appliances' electric consumption amount was conducted, and the results indicated us overall of 9 % electricity savings [17]. In Canada Ontario, the real time feedback from in home displays project was conducted in the summer of 2004 and at the end that study gave us average 6.5 % reduction in the electricity [18]. In Canada British Columbia, with the participation of 200 customers total over 18 month period between 2005 and 2007, overall of 18 % electricity savings was calculated [19]. In California San Diego with the participation of 300 volunteers in San Diego Gas & Electric Company with using in home displays, total reduction of 13 % was found [20]. In another study which was conducted with the participation of Oslo Energi costumers in Norway with the use of more informative billing system 10 % of total electricity savings was found (Wilhite and Ling, 1995). With the use of real time direct feedback method Hayes and Cone [21] carried feedback study in 80 person student housing complex, they found % 18 of total electricity savings. Feedback study which was carried in Des Moines, Iowa lasted for 106 days, and at the end of the study total savings of 16 % was calculated (Palmer et. al,1977). With using in home displays with the time period of 11 months in Carrboro, N.C. with the participation of 25 single family houses, 12 % less electricity consumption was occurred comparing to previous same time period. Other study only use small in home displays in the location of kitchen with the time period of 3 months, from July to September and the total reduction in the electricity consumption amount accounted for 11 % (Seligman and Darley, 1977). Comparing the direct real time feedback methods with indirect feedback methods another study was conducted and electricity saving ratio between 5 – 15 % was found for direct real time feedback application and 10 % was found for indirect feedback application (Darby, 2006).

### 2.3.2 Saving Results Due to Behavioral and Psychological Effects

Osbaldiston and Sheldon [22] indicated that people who feel more responsibility with preserving the environmental sustainability and who feel more enjoy when doing electricity savings action would perform better than other people who couldn't follow their goals and leave the study uncompleted. Goal setting and reward strategies have behavioral and psychological changing effects as we stated before. With the rewarding strategy in which the consumers got paid if any electricity savings would be occurred in six week of period in Lexington, the total of 33 % reduction in total electricity was found (Winett et. al, 1976). In another study, goal setting strategy contributed to 13 % overall electricity consumption reduction (Becker, 1978). Harkins and Lowe [23] conducted study in order to see goal setting effect on electricity savings and their result was average 15 % of total electricity reduction. Another study with goal setting approached was conducted by McCalley and Midden in Netherlands with the participation of 100 person, and the total result in electricity saving was around 20 % [24]. In Canada another feedback study which depended on reward strategy gave the 12.9 % electricity saving result [25].

## 3. Method

### 3.1 General Information

Five households in Ankara, Turkey took part in our study. This was a real time advanced feedback type of study in the appliance level, so consumers were able to see disaggregated electrical consumption information for each individual appliance. We decided on using real time advanced feedback type with the appliance level in our study, because among all other types of feedback advanced real time with appliance level is the best method that cause in electrical energy savings (Darby, 2006). After one month we had also added LCDs to our feedback systems in the houses in which participants can saw and analyze their consumptions with historical ones, and also we wanted to realize the total electricity saving effect of adding of bigger central LCD display to feedback system. As we stated before, the most effective type of feedback system's display should be chose with small in home displays which are in conjuction with central bigger LCD displays (Wood et. al, 2006).

We planned to continue our feedback project with minimum 3 months up to 5 months max. The reason behind that we wanted to see overall short time feedback effect which was accepted as most effective feedback period, it was accepted that the shorter the feedback time the more effective feedback effect on total electricity consumption (Van Raaiij and Verhallen, 1983). Also we gave in literature before, it was accepted that the time need for changing of behaviors into new habits is minimum three months (Darby, 2006). Because of participants' personal reasons, our period of applying feedback technology products in houses differed, total duration of project in house # 1 was 5 months, total duration in house #4 was 4 months and 10 days, total duration of project in house #5 was 3 months and 7 days, total duration of project in house 2 was 4 months and total duration in house 3 was 3 months 12 days. In the beginning participants from seven houses accepted to be in the study, after the start of our project house # 7 indicated they didn't want to continue with the study after 1 month and the problem with the number # 6 was; they joined the study with 4.5 months but they couldn't provide their previous years' electrical consumptions. So we surveyed all of the seven houses before starting to our feedback study with questionnaire in order to obtain the energy consumption knowledge of participants. We didn't give any electricity saving suggestions to participants, the reason of that was being sure about avoiding Hawthorne effect [26], in which participants would behave differently if they got any information and thus we couldn't be able to get the real results. Thus, we hadn't made any contact with participants after the experiment was under way. Since house # 6 couldn't provide their consumptions, we could only compared overall feedback effect with the previous month before we started our project.

For finding the total feedback effect on reducing the electrical usages on households, electrical energy consumption amounts of five household compared with the previous year consumption amounts with same time periods. Electrical consumption amounts for the previous year for each households were obtained from the Enerjisa Company. The total percentage feedback effect (TPF) measure was calculated as follows:

$$TPF = \frac{\text{total measured consumption (our study)} - \text{total historical consumption}}{\text{total historical consumption}}$$

The negative result of TPF indicated that with positive feedback effect total electrical consumption was reduced, contrary positive result of TPF indicated that participants consumed more than previous years.

## 3.2 Survey

We made a survey in which participants have to answer 37 different questions. We aimed to reveal participants' some thoughts in electrical consuming activities, possible electricity saving approaches of households, characteristic features of consumers and factors that may have effect on participants for investing in new technologies. Questionnaire was examined under the following categories.

(1)     Socio demographic features of households: we tried to make any connection between electrical usage patterns of participants and socio demographic features like age, education level, population of house, occupations and income level of houses.

 (2)     In home activities: in that category we tried to learn whether the participants have specific hobbies, activities they do in their houses, we asked total estimated usage time of appliances like TVs, game consoles, musical system.

(3)     Usage-patterns for appliances: we asked the total estimated usage frequencies of specific appliances like fridges, televisions, laptops, washing machines, dish washers etc.

(4)     Major consumers: in that part of survey, we tried to evaluate their assumptions about electrical consumption amounts of appliances and tried to draw conclusion with total knowledge of participants over electrical appliances consumption levels, we asked them please give points between $1 - 5$, if you think that specific appliance consumes the highest give point 5.

(5)     Electricity consumption awareness: we investigated participants' awareness level over electric consumption amounts, whether they used to control electric bills, compare the total electricity consumption amounts in the bills with previous months, and tried to find out what do they think about themselves, are they economizer or extravagant.

In the demographic part of survey, we addressed some questions to participants with the intention of trying to create any relationship between socio demographic features of households (see Table 1) and the total consumption levels and also possible further interest to energy efficient behaviors that let any saving in the total electrical usage.

All of the houses have high income level, financial situation of households weren't significantly different from each other, beyond all houses; house 5 and house 6 income levels were the highest. House 2 was an office, the population could differ with time but office had 10 employees who work full time. In house 7 there was a family with one small child who aged 5, except house 7 other people in households were adults. All of the houses were owner - occupied, there weren't any rentals in our study. Except house 3 and house 4 all of the houses have floors number of more than one, house 6 have most floor number of 4.

In electricity consumption awareness part, we wanted participants to evaluate their own electrical consumption behaviors up to start point of our study. We wanted to learn whether they thought they had demonstrated economizer, spender or normal consumer level up till now. Also we wanted to learn what they thought about possible future reduction level in total electricity savings due to adapting of new positive electricity savings behavior and we addressed that question to participants: "In which rates of electricity savings that you will make if you change your consumption behavior?" (See Table 2). We saw bigger possible future savings percentage amounts responds which belonged to participants who thought their total consumption level was too much. Participants who thought their total electric consumption level slightly exceeded the desired electrical consumption level, they believed that they can made $3 - 10$ % possible future savings. As seen from the table neither of the participants didn't say their electrical consumption level was normal indicated that all of the participants were aware of excessive use of electrical energy in their houses and with the intention of saving some amount they wanted to participate in our study.

**Table 1.** Demographic features of participants and household

|  | Gender | | Tenure | | Age | | | | Income level (Turkish Liras) | | | | Education Level | | | # of floors |
|---|---|---|---|---|---|---|---|---|---|---|---|---|---|---|---|---|
|  | M | F | OO | R | <17 | 17-30 | 30-45 | 45> | 8k-9k | 9k-10k | 10k-15k | 15k> | HS | BSc | MSc |  |
| House 1 | 2 | 1 | OO |  |  | 1 |  | 2 | X |  |  |  |  | 2 | 1 | 2 |
| House 2 | 7 | 3 | OO |  |  | 1 | 6 | 3 |  |  | X |  |  | 9 | 1 | 2 |
| House 3 | 2 | 1 | OO |  |  | 1 |  | 2 |  | X |  |  | 1 | 1 | 1 | 1 |
| House 4 | 1 | 1 | OO |  |  |  |  | 2 |  | X |  |  |  | 2 |  | 1 |
| House 5 | 2 | 2 | OO |  | 2 |  |  | 2 |  |  |  | X |  | 3 | 1 | 3 |
| House 6 | 2 | 1 | OO |  |  | 1 |  | 2 |  |  |  | X |  | 3 |  | 4 |
| House 7 | 2 | 1 | OO |  |  | 1 |  | 2 |  |  |  | X |  | 2 |  | 3 |

M = male, F = female; OO = owner-occupied, R = rental; HS = high school, BSc = bachelor of science, MSc = master of science

**Table 2.** General thoughts about consumption levels and possible future savings rate

|  | General thought for consumption levels up to start point of our study | Possible future savings rate |
|---|---|---|
| House 1 | "My consumption had been slightly more than the level it should be" | 3 – 5 % |
| House 2 | "My consumption level was too much more than the level it should be" | 10 – 15 % |
| House 3 | "My consumption level was too much more than the level it should be" | 10 – 15 % |
| House 4 | "My consumption had been slightly more than the level it should be" | 5 – 10 % |
| House 5 | "My consumption had been slightly more than the level it should be" | 3 – 5 % |
| House 6 | "My consumption had been slightly more than the level it should be" | 5 – 10 % |
| House 7 | "My consumption had been slightly more than the level it should be" | 5 – 10 % |

When filling the form part of the three highest electrical energy using devices in households, choosing of multiple appliances was solicited, participants could choose as much number as they wanted for the number 1, 2 and 3 nominees of the highest electrical energy consuming devices. Respondents were asked to indicate which three domestic electrical appliances they thought contributed most to the electricity bills in their household (see Table 3). Fridges were identified by the highest electrical energy consuming device with the ratio of 18.80 %, nominee for the second highest electrical energy consuming device was selected as lighting systems in the houses with the ratio of 26.70 % and washing machines were chosen for the third highest electrical energy consumption device in households with the ratio of 20.00 %.

**Table 3.** Appliances chosen by the participants as the three highest electrical energy consuming devices in households.

| Respondents' perception | Appliance's choosing proportion by respondents (%) |
|---|---|
| Highest electricity user | |
| Fridge | 18.80 |
| Dish washer | 6.24 |
| TVs | 12.50 |
| Cooker | 12.50 |
| Computers & Laptops | 12.50 |
| Air conditioner | 6.24 |
| Lighting systems | 6.24 |
| Iron | 12.50 |
| Clothes dryer | 6.24 |
| Washing machine | 6.24 |
| Second highest electricity user | |
| Dish washer | 13.33 |
| TVs | 13.33 |
| Cooker | 13.33 |
| Computers & Laptops | 6.66 |
| Lighting systems | 26.70 |
| Iron | 13.33 |
| Clothes dryer | 6.66 |
| Washing machine | 6.66 |
| Third highest electricity user | |
| Dish washer | 15.00 |
| TVs | 10.00 |
| Cooker | 10.00 |
| Iron | 10.00 |
| Clothes dryer | 10.00 |
| Washing machine | 20.00 |
| Air conditioner | 5.00 |
| Musical system | 10.00 |
| Fridge | 5.00 |
| Game console | 5.00 |

**3.3 Design and Materials**

Classification of feedback technologies over different type of categories was needed because feedback technology product's efficiency depends on different characteristic features. In literature there was a need for detailed categorization of the feedback technology products. Without omitting technical and psychological effects of different feedback technology products, extensive literature research had been done. In the study of Karlin, classification of 196 various feedback devices had been executed via empirical method. In that study, taxonomy was derived based on the different characteristics features; information platform, management platform, appliance monitor, load monitor, grid display, sensor display, networked sensor, closed management network and open management network [27].

In our study, we used electrical feedback product called Geo II in home display which is manufactured by the company Green Energy Options. With using of Geo II in home display product, we were able to see real time total electricity consumption of individual households. Also by using the smart plugs, which came with Geo II in home display, we were able to analyze the disaggregated total consumption amount for each electrical appliance we selected for recording. In one total system, one Geo II in home display product was able to connect to maximum 6 units of smart plugs, we were only able to reveal the consumption profiles and record the continuous electrical energy consumptions of 6 electrical consuming devices. Smart plugs also provided remote on off control from internet via signals, participants were able to open or close the appliances when they wanted. In addition to these features, company provided one year free web database page in which participants could see the real time total electrical consumptions, historical consumption amounts and also they could see all of the selected appliances' total electrical consumptions and related currency amounts of consumptions.

Fig. 1 shows the configuration of electrical consumption feedback technology system that we set up in the households. The whole system comprised of measuring, transmitting, monitoring components and also web service. The measurement of the current total electricity was quantified by clipping sensors which were clipped to main electrical fuse box 3 phase live cables. At the other end of clipping sensors, cable sockets exist, those cable sockets was putted into transmitters. Transmitters sent total real time electrical consumption information to in home display and making the consumption amount visual. Geo II in home display devices sent consumption information to web bridge device via wireless signals. Connecting the internet bridge with an Ethernet cable to modem that we used in households, web bridge gain the ability of sending the total electrical consumption information to own web energy service database. Disaggregated feedback information maintained by using the smart plugs, smart plugs was paired with Geo II in home displays. Smart plugs sent the consumption amount of plugged electrical energy consuming device via wireless signals to in home displays. Internet bridges also sent that individual appliances' consumption information to energy web service. Monitoring components include bigger LCDs in which the online web energy service of our product was continuously open and participants could analyze and see the consumptions amount when they want.

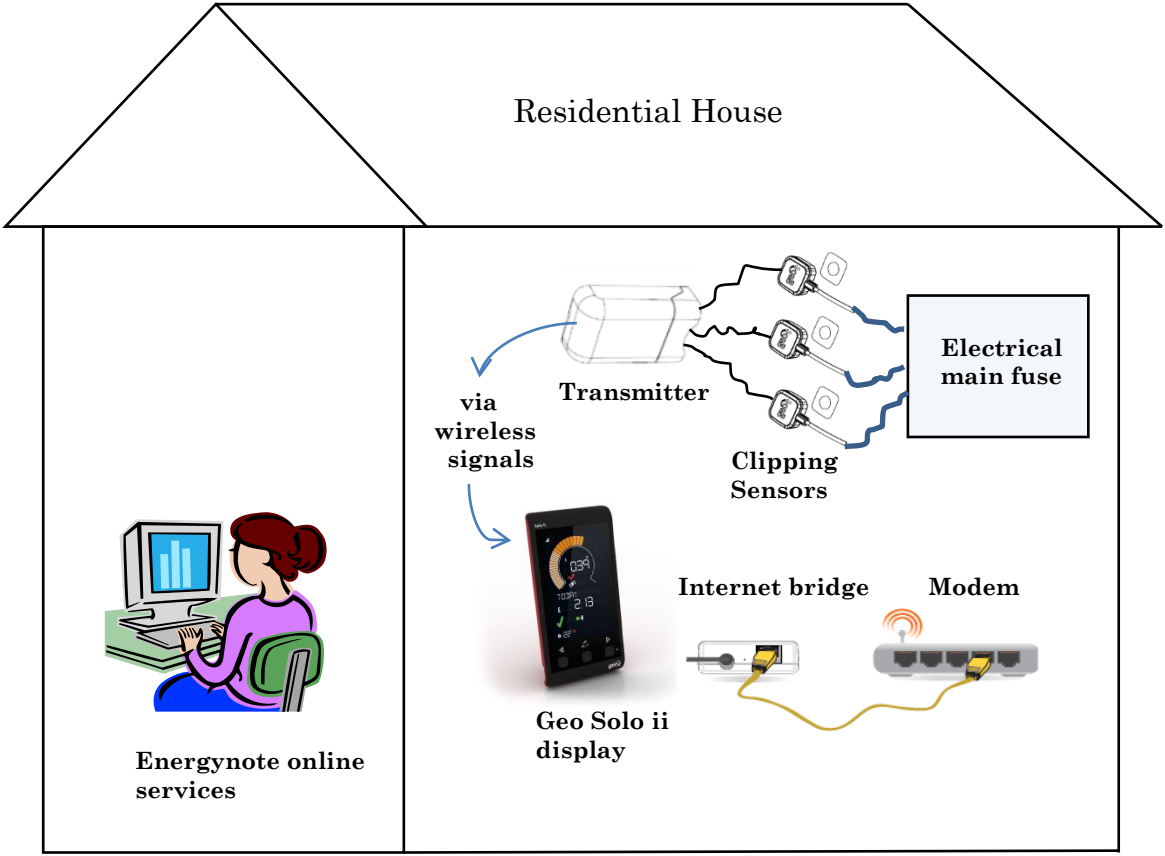

Figure 1. Configuration of our electrical consumption feedback technology product

### 3.4 Goal Setting

With the intention of avoiding any inducement that could have effects on participants, we simply gave over the determining the goal to consumers. When we asked to determine their budget goals, they said they need to check the previous months' electrical bills and some of them weren't sure about whether they keep their electrical bills or not, in the final the budget goals for all houses 1-5 were; 160, 300, 90,90, 150 Turkish Liras correspondingly.

### 4. Results

#### 4.1. Survey

We made pre survey before to start our study, fridges were nominated for the most consuming appliance with the 18.80 % ratio, total expenditure of the lighting systems were nominated for the second most consuming system and for the third most consuming electrical appliance would be washing machine with the ratio of 20.0 % (Table 3). We gauged the total kilowatt hour usages of individual appliance in 6 households. Beyond any doubt, lighting systems have big share in total monthly electrical usage, however the feedback device only could gauge individual appliances due to we made analysis over individual devices. During the time period of our study, three most electrical energy consuming appliances in each household and related ratio amounts to total electrical usage were shown in Table 4. The most electrical energy consuming device was fridge with the total monthly amount of 77.5 kWh which accounted for the 30.59 of the total electrical usage for the house 3 in the April, the nominee for the second most electrical energy usage was printer with the total monthly amount of 49.6 kWh however we couldn't accept as second because it was only located in house 2 which was an office so the second most electrical energy using device was dish washer with the total monthly amount of 43.1 kWh; the total amount accounted for 10.5 % of the total electrical energy usage in house 1 in May, the third most electrical energy using device was electrical oven/cooker with the total monthly amount of 41.6 kWh; which accounted for 7.68 % of the total electrical energy usage in house 1 in April. Comparing with the pre survey, we could say 18.80 % of the participants have the right idea for the most electrical energy consuming device and for the third electrical energy using consumption devices they couldn't guess right. Over the long run, we suggest further enlightenment programs with the electrical usage subject will should transfer into practice in the schools (primary schools to high schools) which oblige the participation of children and guardianships. We believed that kind of informing may cause huge electrical usage savings in total in country wide.

#### 4.2. Goal Setting

Participants asked to determine their own budget goal before start our studies in their homes. It was seen that participants determined their budget goal with only considering 1 or 2 months before. That kind of approach could be accepted inherently however, it was not the exact way of approach when deciding the right goal budget setting. In Table 5 determined budget goals for all houses; house 1 to house 5, and their previous 3 months electrical consumption monetary values can be seen. In the pre study survey, house 1 participants thought their total monthly consumption was slightly more than the level it should be, their budget goal was 8.05 % and 19.56% lesser than to December 2015 and January 2015 total consumption correspondingly. In pre survey house 2 participants stated that their consumption was too much more than the normal level, during the study they decided that their budget goal level was 300 Turkish Liras and that level was 17.68 % lesser than the January 2015, one month before, that situation clarified that house 2 participants only took last month consumption into account. House 3 participants also stated their total consumption amount was too much more, and their budget goal was 9.82 % and 28.15 % lesser than December 2015 and January 2015 correspondingly. House 4 participants thought their consumption level was slightly more than the normal level and in the final they decided their budget goal should be 90 Turkish Liras which was 20.64 % and 9.89 % lesser than the December 2015 and January 2015 correspondingly. Like house 4 house 5 participants thought their total consumption amount was slightly more than the desired level, they decided their goal budget should be 150 Turkish Liras, which accounted for 31.17 % and 32.28 % lesser than to December 2015 and January 2015 correspondingly. It was revealed that determining the budget goal with only looking to 1 or 2 months before can't be preferred way of approach due to that kind of approach clearly mislead the participants and could led consumers to consume more than the desired level. We suggested finding the previous one year's average monthly monetary value, in which we calculated the total electrical cost value with taking the total monthly electrical consumption amount and relevant month's electrical tariff price and made some calculations, and after finding the resulted total cost for all months, we took the average. In Table 6 average monthly monetary value of electrical consumption and comparing with the participants budget goal can be seen.

**Table 4.** Individual appliances total monthly electrical consumption

|         | # 1 | # 2 | # 3 |
|---------|-----|-----|-----|
| House 1 | TV: 44.0 kWh (9.55 %), June | Dish Washer: 43.1 kWh (10.5%) (10.5 %), May | Electrical Oven: 41.6 kWh (7.68 %), April |
| House 2 | Printer: 49.6 kWh (9.58 %), April | PC System 1: 25.7 kWh (4.96 %), April | PC System 2: 25.6 kWh (7.97 %), May |
| House 3 | Fridge: 77.5 kWh (30.59 %), April | Dish Washer: 36.1 kWh (16.63 %), May | TV: 16.5 kWh (7.60 %), May |
| House 4 | TV: 46.8 kWh (24.18 %), May | Fridge: 46.2 kWh (17.61 %), April | Freezer: 35 kWh (16.21 %), June |
| House 5 | Fridge: 64.0 kWh (12.02 %), June | Dish Washer: 31.2 kWh (5.86 %), June | Water Heater: 27.1 kWh (7.37 %), May |
| House 6 | Fridge: 58.3 kWh (9.3 %), June | PC System: 23.6 kWh (3.8 %), June | Iron: 14.7 kWh (2.34 %), June |

As seen from the Table 6, budget goal of house 1 was 6.18 % lesser than the average electrical cost of previous one year. Budget goal was 16.54 % lesser than the average electrical cost of house 3 in previous year. House 4 and house 5's percentage differences that states the percentage difference with the budget goals and average electrical cost were 8.73 % and 10.67 % correspondingly. There was a misleading error due to participants in house 2 determined their budget goal only looking to one month before in which they had extensive use of electrical with the result of using heaters in office rooms, and their budget goal was 64.66 % higher than the average electrical cost. Determining the rational budget goals was so important in adapting the electrical savings behavior activity.

**Table 5.** Household's previous 3 months electrical consumption monetary values and determined budget goal values (Turkish Liras)

|         | November 2015 | December 2015 | January 2015 | Budget Goal |
|---------|---------------|---------------|--------------|-------------|
| House 1 | 156.719 TL    | 173.990 TL    | 198.904 TL   | 160 TL      |
| House 2 | 160.481 TL    | 296.428 TL    | 364.460 TL   | 300 TL      |
| House 3 | 76.568 TL     | 99.807 TL     | 125.261 TL   | 90 TL       |
| House 4 | 82.661 TL     | 113.405 TL    | 99.886 TL    | 90 TL       |
| House 5 | 155.532 TL    | 217.936 TL    | 221.509 TL   | 150 TL      |

**Table 6.** Average monetary values of electrical consumption (Turkish Liras)

|         | Average electrical cost | Budget Goal | Percentage Difference (%)      |
|---------|-------------------------|-------------|--------------------------------|
| House 1 | 170.54 TL               | 160 TL      | 6.18                           |
| House 2 | 182.19 TL               | 300 TL      | Misleading error was occurred. |
| House 3 | 107.83 TL               | 90 TL       | 16.54                          |
| House 4 | 98.61 TL                | 90 TL       | 8.73                           |
| House 5 | 167.92 TL               | 150 TL      | 10.67                          |

If we determined the budget goal as so low that impossible to reach, the motivation of consumers probably will be loosed, if we determined the budget goal was so high that is almost the same as the average monthly monetary values of electrical consumption, we couldn't make any changes in consumers' behaviors. House 4, house 5 and house 1 participants thought their consumption level was slightly more than the normal level, house 2 and house 3 participants thought their consumption level was too much more than the normal level as we stated before in the Table 2. So we suggested specified level of percentages in electrical consumption reducing: in homes whom participants think their consumption level was slightly more than the normal level we suggested the total reducing in budget goal with the ratios between 1 – 15 %, in homes whom consumers think that their consumption level was too much more than the desired level we suggest budget goal that make reducing of 15 – 20 % in total cost.

### 4.3. Results of Feedback Study

#### 4.3.1. General Results of Feedback Study

Of the 7 households that actually started the experiment, 6 completed it. However, we made feedback analysis over the 5 houses due to house 6 couldn't provide its' historical electrical consumption amounts. House 7 quitted with personal reasons specifically we couldn't have found the time for interest. The feedback results of house 1 to house 5 can be seen in Table 7. Among our five households we couldn't get any electricity savings in house 1 and house 2, on the contrary 361.653 kWh and 483.307 kWh increase in total electrical consumption amount was occurred in the houses 1 and 2 correspondingly. Before the study, participants in the house 1 indicated that there might be problem on total consumption in their refrigerators and the television in the kitchen. With that information we wanted to apply appliance level feedback and wanted to record the electrical consumptions of refrigerators, however we couldn't gauge because the refrigerators were embedded type and we couldn't reached the receptacles behind. Except that inconvenience, we put our smart plug into kitchen television and saw that the total electrical consumption of television in the kitchen in the June was nearly three times of television in house 3 in May, and also was 2.2 kWh lesser than the total electrical consumption of fridge in the house 4 in the month April. That specific electrical appliances may be the reason of we couldn't get any electricity savings in house 1, we believed refrigerators and television in the kitchen might use more electrical energy due to historical electrical problem that was occurred in devices, maybe because of that we couldn't see any electricity savings because the total saving effect of feedback devices might be covered up and there is obvious need for change and investing and replacing with new ones.

House 2 was a running office and we couldn't get any electricity saving effect in there too. We believed main reason of that was population changing reality. Population in the office differed frequently and because of that the total consumption might be increased. Such changes have such impact on energy requirements that no conclusions can be drawn on the behavioral change of the households. Also, running the heaters in the winter could other reason of increase in the total electricity. Before the study, they declared that they used the heaters nearly 3 months and because of that we put an appliance level feedback product, smart plug to heater however increase in the general temperature after start of March, they didn't need to use heaters so we couldn't make any analysis, however it is also obvious that they need to invest in insulations or buy more efficient air conditioners for heating in the winter. One of the other reason can be employees didn't show any importance to our feedback study. Because they only worked for the company and they were not in the responsibility to pay the monthly total electrical price cost. It was seen that one of the key factor in order to feedback technology can result in electrical energy reducing is ownership level of participants.

We applied our feedback study in house 3 from 18 February to 31 May, at the end when we compared with the total electricity amount in this period with the previous year we got the 13.21 % feedback effect. In the pre survey the participants declared that they consumed too much than the normal level. It showed that awareness level of the historical consumption and motivation for making reduction in total electricity savings are very important factors for reducing the total electrical consumption and making some savings with the feedback technology. House 4 and house 5's feedback effects were 3.42 % and 1.59 % correspondingly. They thought their consumption level was slightly more than the desired level.

The feedback effects of house 3 was the best, feedback effect of house 4 was became second and the feedback effect of house 5 with 1.59 % became third. House 3 and house 4 are single floor houses, whereas house 5 is 3 triplex and house 1 and house 2 are two storied. When getting more effective electricity savings results with the applying of feedback technologies, we will suggest applying feedback technologies studies in the single floor houses. Realization rate of in home displays and also LCDs in single floor houses is higher two storied or triplex houses, due to consumers can change their consumption behaviors and can adapt the new behaviors easier in those houses.

It was also important to analyze in which time period participants were most likely to change their behaviors, in which time period feedback effect was the highest. In Turkey consumer can select mono chromic electrical tariff prices or multi time electrical tariff prices. In multi time electrical tariff pricing system, day time period is between 06:00 am – 17:00 pm, peak time period is between 17:00 pm – 22:00 pm and night time period is between 22:00 pm – 06:00 am. Electrical counters measured the total electrical energy consumption in each period. In our study feedback technologies let to reduce the total electricity savings in house 3, house 4 and house 5. However, we could only take the total electricity consumptions, we couldn't see exact consumption amounts of day, peak and night time periods of house 4 and house 5 because they connected electrical counter like nine months ago. We could only make feedback effect analyze in home 3 with different time periods and comparing them. Feedback results that belong to different time periods could be seen in Table 8. The biggest saving in electricity was between 06:00 am – 17:00 pm with the ratio of % 16.40, 14.78 % and 5.84 % electricity savings belonged to peak and night time period correspondingly. However with the results from one house it would be impossible to make any general inferences, we suggested to researchers to concentrate on changing the consumers' behaviors in the day and peak time periods. When adapting the new consuming behaviors or investing in new energy efficient technologies, the specific devices or behaviors in the time periods of day and peak should be esteemed more.

**Table 7.** Feedback Effect

|  | Total Electrical Cons. in 2015 (kWh) | Total Electrical Cons. in 2016 (kWh) | Difference (kWh) Inc. or Dec. | Feedback Result ( % ) |
|---|---|---|---|---|
| House 1 Per.: 16 Feb. – 17 July | 2036.269 | 2397.922 | 361.653 (Increased) ( 17.76 % ) | 0 – No effect |
| House 2 Per.: 18 Feb. – 17 June | 1257.258 | 1740.555 | 483.307 (Increased) ( 38.44 % ) | 0 – No effect |
| House 3 Per.: 18 Feb. – 31 May | 952.008 | 826.204 | 125.804 (Decreased) | 13.21 |
| House 4 Per.: 20 Feb. – 30 June | 1016.715 | 981.892 | 34.824 (Decreased) | 3.42 |
| House 5 Per.: 23 Feb. – 31 May | 1210.183 | 1190.928 | 19.255 (Decreased) | 1.59 |

Cons. = Consumption, Inc. = Increase, Dec. = Decrease, Per. = Period, Feb. = February

**Table 8.** House 3 Feedback Effect Comparing within different time periods

| | Feedback Result ( % ) | | |
|---|---|---|---|
| Time Period: 18 February – 31 May | | | |
| | Day (06:00 – 17:00 pm) | Peak (17:00 pm – 22:00 pm) | Night (22:00 pm - 06:00 am) |
| House 3 | % 16.40 | % 14.78 | % 5.84 |

### 4.3.2. Comparing the Feedback Effects

Our other motivation in was analyzing the effect of adding central LCD to houses together with the appliance level devices and comparing the feedback effect when only in home displays were set up in the houses. With that kind of approach we set up our feedback devices, in home displays to all houses; house 1 to 5, than after one month we added LCDs and also our smart plugs that gave the capability of measuring and analyzing appliance level feedback. As we stated in previous part of our study, we got electricity savings in home 3, 4 and 5 so we analyzed the effect of adding central LCDs and adding appliance level devices in home 3, 4 and 5.

We set up the in home display in home 3 and started our study in 18 February, till to 17 March only in home display was showing the electrical consumption information. During 18 March - 31 May participants had chance to reach their appliance level analysis and also had chance to see results from bigger central LCD. In home 4, in home display was set up and started to measure in 20 February, till to 20 March only in home display was displaying the electrical consumption information. During 21 March – 30 June, consumers could reach their appliance level analysis and had chance to see the consumption information and analysis on bigger central LCD. In home 5, in home display was set up and started to measure in 23 February, till to 23 March only in home display was showing the electrical consumption information. During 24 March – 31 May, participants could reach their appliance level analysis and had chance to see the consumption information and analysis on bigger central LCD. Comparing results can be seen in Table 9.

**Table 9.** Comparing the feedback effects

|  | Feedback Result ( % ) |
|---|---|
| House 3 | |
| I.H.D: 18 Feb. – 18 March | 5.53 |
| A.L. & LCD: 19 March – 31 May | 16.19 |
| House 4 | |
| I.H.D: 20 Feb. – 20 March | 0.96 |
| A.L. & LCD: 21 March – 30 June | 4.16 |
| House 5 | |
| I.H.D: 23 Feb. – 23 March | 3.34 |
| A.L. & LCD: 24 March – 31 May | 0.90 |

I.H.D = In home display, A.L. = Appliance level

In house 3 and house 4 when applying appliance level and central LCD together, we got bigger feedback effects which caused bigger electricity savings. In house 3 when only in home display gave the feedback information to consumers, 5.53 % electricity savings to to previous year had calculated, however when applying the appliance level feedback with central LCD feedback effect was 16.19 %. Like house 3, in house 4 when applying the appliance level of feedback with central LCD together 4.16 % electricity savings to previous year was obtained. In house 5, the feedback effect caused by in home display was bigger than the complete feedback system. The situation was different in house 5, in general they made electricity savings with the ratio of 1.59 % and we fathomed out that participants in house 5 may began to give up the electricity savings behaviors and may be they would consume more, we reached that kind of thinking by comparing the appliance level feedback effect together with central LCD with only in home display feedback effect. Results of house 3 and house 4 supported our approach. Besides, as we stated before, the most effective type of feedback system's display should be chose with small in home displays which are applied together with central bigger LCD displays (Wood et. al, 2006).

## 5. Conclusions

Feedback technologies have great potential on reducing the total electricity consumption in households. Technological features of feedback devices and socio demographic features of households play key role on effectiveness of feedback technology products. Real time advanced appliance level feedback technologies which give the capability of analyzing the consumption in real time by looking into in home displays together with bigger central LCDs have the highest potential on reducing the total electricity consumption. In house 3 and house 4 we saw that applying the feedback technology product in appliance level together with central bigger LCD gave higher electricity savings. Floor number of houses is also another important factor, we don't suggest application feedback technologies in more than two floors houses, if it is desired we recommended putting central LCDs to every floor. In addition, in our study applying the feedback technologies product in offices didn't work well, we only suggest applying in houses. Determining the goal budget issue is so important aspect when applying the feedback technologies. If it is desired to make reduction as much as possible, we suggest determining the goal budget by taking the previous year's average monthly electrical consumption cost and take the goal budget minimum 15 % under this cost. House income level is another important issue for success of feedback technologies, if households' income level is so high (higher than 15000 Turkish Liras monthly) they may not give the necessary caution to feedback technology product to become successful and make some electricity savings results. We believe influencing the consuming behaviors of participants whose monthly income is maximum 10000 Turkish Liras will be more effective. We suggest making feedback technology studies that will based on time intervals that concentrating on time periods in which comparing the consumption amount of participants and try to analyze consumers behaviors in that different time periods and try to reveal in which time period it is more possible to motivate the consumers to change their behaviors and it is more possible to make bigger savings.